\documentclass[journal]{IEEEtran}

\usepackage{xcolor,soul,framed} %
\usepackage{booktabs}
\usepackage{siunitx}
\colorlet{shadecolor}{yellow}
\usepackage[pdftex]{graphicx}
\graphicspath{{../pdf/}{../jpeg/}}
\DeclareGraphicsExtensions{.pdf,.jpeg,.png}

\usepackage{changes}
\definechangesauthor[name={R0}, color=orange]{R0}
\definechangesauthor[name={R1}, color=blue]{R1}
\definechangesauthor[name={R2}, color=red]{R2}
\definechangesauthor[name={R3}, color=green]{R3}

\usepackage[cmex10]{amsmath}
\usepackage{amsfonts}
\usepackage{array}
\usepackage{mdwmath}
\usepackage{mdwtab}
\usepackage{eqparbox}
\usepackage{url}
\usepackage{multicol, blindtext}
\usepackage[caption=false]{subfig}
\usepackage{caption}
\usepackage{comment}
\usepackage[noadjust]{cite}

\usepackage[switch,columnwise]{lineno}

\usepackage{tikz}
\usetikzlibrary{shapes,arrows,positioning,calc}
\usetikzlibrary{arrows.meta, calc, chains, positioning}
\usetikzlibrary{
	decorations.pathreplacing,
	shapes.symbols
}

\definecolor{OliveGreen}{rgb}{0,0.6,0}

\begin{document}

\bstctlcite{IEEEexample:BSTcontrol}   	
\title{Soft-Demapping for Short Reach Optical Communication: A Comparison of Deep Neural Networks and Volterra Series}    	
  \author{Maximilian~Schaedler,~\IEEEmembership{Student Member,~IEEE,}
      Georg B{\"o}cherer,~\IEEEmembership{Member,~IEEE,}\\
      Stephan Pachnicke,~\IEEEmembership{Senior Member,~IEEE.}

 \thanks{Manuscript received October 22, 2020; revised December 27, 2020; accepted January 19, 2021. Date of publication XXXX, 2021; date of current version January 26, 2021. The author(s) received no specific funding for this work.}
 \thanks{M. Schaedler is with Kiel University (CAU), Chair of Communications, Kaiserstr. 2, 24143 Kiel, Germany and with Huawei Munich Research Center, Riesstr. 25, 80992 Munich, Germany (e-mail: maximilian.schaedler@huawei.com).}
 \thanks{G. B{\"o}cherer is with Huawei Munich Research Center, Riesstr. 25, 80992 Munich, Germany (e-mail: georg.bocherer@huawei.com).}%
 \thanks{S. Pachnicke is with Kiel University (CAU), Chair of Communications, Kaiserstr. 2, 24143 Kiel, Germany (e-mail: stephan.pachnicke@tf.uni-kiel.de).}}

\maketitle

\begin{abstract}
In optical fiber communication, optical and electrical components introduce nonlinearities, which require effective compensation to attain highest data rates. In particular, in short reach communication, components are the dominant source of nonlinearities. Volterra series are a popular countermeasure for receiver-side equalization of nonlinear component impairments and their memory effects.  However, Volterra equalizer architectures are generally very complex. 

This article investigates soft deep neural network (DNN) architectures as an alternative for nonlinear equalization and soft-decision demapping. On coherent 92GBd dual polarization 64QAM back-to-back measurements performance and complexity is experimentally evaluated. The proposed bit-wise soft DNN equalizer (SDNNE) is compared to a 5th order Volterra equalizer at a 15\% overhead forward error correction (FEC) limit. At equal performance, the computational complexity is reduced by 65\%. At equal complexity, the performance is improved by 0.35~dB gain in optical signal-to-noise-ratio (OSNR).
\end{abstract}


\IEEEpeerreviewmaketitle

\section{Introduction}

\IEEEPARstart{I}{n} modern communication systems, soft decision forward error correction (FEC) and high-order quadrature amplitude modulation (QAM) schemes are key technologies for realizing high spectral efficiencies (SE)~\cite{alvarado2015replacing}. Whereas the first optical communication systems relied on hard-decision (HD)-FEC, nowadays, modern systems use soft-decision (SD)-FEC. Current digital coherent systems typically employ bit-interleaved coded modulation (BICM)~\cite{caire1998bit} and rely upon bit-wise (BW) decoders that separate signal recovery and FEC. A key component of a SD-FEC BICM system is the soft demapper, which computes the soft bits in the form of the so-called $L$-values, expressed as loglikelihood ratios (LLRs)~\cite[Chap.~16]{lin2004error}. The $L$-values feed the following error correction decoder and, thus, affect directly with their quality the overall system performance. Classical approaches to evaluate the $L$-values, as the analytical computation of the a-posteriori LLR~\cite[Sec. 3.3]{szczecinski2015bit} or the less complex max-log approximation (MLA)~\cite[Sec. 3.3.3]{szczecinski2015bit}, assume ideal channel compensation of the channel impairments and thus additive white Gaussian noise (AWGN). However, an optical communication system comprises even in back-to-back (BtB) configuration residual impairments from a large number of optical/electrical (O/E) components with nonlinear memory effects that, in practice, cannot be completely compensated. Their limitations on the achievable capacity are aggravated by measures towards higher data rates, such as symbol rate increase or the shift to higher order QAM. Therefore, computing the $L$-values under the AWGN assumption implies a performance penalty. 

Nonlinear compensation for O/E components is not yet a standard feature in today’s optical transceivers, but will inevitably become a key element of digital signal processing (DSP) to keep up with ever increasing data rates. In recent years, machine learning (ML) methods, especially deep neural networks (DNN), demonstrated excellent performance gains and complexity reduction approaches in various applications. In the optical communication community, DNNs have been applied in several publications on short-reach IM/DD systems to characteristic O/E components~\cite{rios2017experimental,reza2018nonlinear,li2018100gbps} as well as on long-haul coherent links to tackle fiber nonlinearities~\cite{koike2018fiber,kamalov2018evolution,zhang2019field,sidelnikov2018equalization}.
In~\cite{koike2018fiber,schaedler2020neural}, where ML is applied for $L$-values computation, the authors have shown that DNNs are a convenient tool to predict probabilities and hence, to learn soft demodulation schemes, where a received symbol has to be classified to its transmitted bits.

This paper builds upon the aforementioned contributions with emphasis on high-bandwidth optical short reach coherent communication systems, where O/E component nonlinearities dominate. Typical use cases include data center interconnects (DCI) with a range of 80–120~km~\cite{bluemm2019single}. The performance and computational complexity of a proposed soft DNN equalizer (SDNNE) is benchmarked against a Volterra nonlinear equalizer (VNLE) accompanied by a soft-demapper on basis of coherent dual polarization (DP) 92GBaud 64QAM back-to-back (BtB) ofﬂine captures. The VNLE is a popular approach, which has proven to be very effective against component nonlinearities~\cite{rezania2016compensation,cartledge2017volterra}. It can be tailored to match any differentiable nonlinear system by choosing a high enough polynomial order and memory depth~\cite{schetzen1976theory}. However, excessive extension for optimal performance can become a significant downside, as the architectural complexity of VNLEs increase exponentially. For a detailed comparison of both approaches, we follow D. I. Soloway \textit{et al.} \cite{soloway1992neural} and extract the linear and nonlinear kernels from the trained SDNNE by Taylor expansion and compare with the Volterra kernels. Furthermore, several SDNNE architectures with different activation functions and representation capacity are compared in terms of performance and complexity.

The remainder of the paper is structured as follows. In Section II, the considered equalizer architectures and their relationship  is introduced. Afterwards, the experimental investigations are shown and analyzed in Section III. Section IV compares the particular complexity. Finally, the paper is concluded in Section~V.

\section{Nonlinear Component Equalizers}
In this section, we introduce the general VNLE and the DNNE alternative. We discuss the principles as well as the relationships between both approaches.

\subsection{Principles of General Volterra Nonlinear Equalizers} \label{pV}
By combining linear convolution and nonlinear power series the VNLE is capable to describe causal as well as non-causal time-invariant nonlinear systems with finite fading memory. With $y(k)$ and $\Tilde{y}(k)$ representing system single-input and single-output, respectively, the $P$-th order non-causal discrete time Volterra series is given by~\cite[Sec.~4.2]{guan2017fpga}
\begin{equation}
\begin{split}
     &\Tilde{y}(k) = \label{eq.:VNLE}    \\ & \sum_{p=1}^{P} \sum_{s_1=-M_p}^{M_p} \dotsb\sum_{s_p=k_{p-1}}^{M_p} h_p(s_1,\dotsb, s_p) \prod_{i=1}^{p} y(k-s_i),
\end{split}
\end{equation}
where $h_p(s_1,\dotsb, s_p)$ denotes the pth-order Volterra Kernel, $M_1$ the symmetric memory length for the linear terms and $M_2$ to $M_p$ the symmetric memory lengths for the nonlinear terms of second order and higher. Generalizing \cite[Table~I]{wettlin2020complexity} to an arbitrary order $p$, the relationship between the memory lengths and the number of equalizer kernels is given by
\begin{equation}
N_p = \frac{1}{p!} \prod_{i=0}^{p-1} (M_p + i).
\end{equation}
Before operations, the kernels have to be identified, i.e., configured upon training data, in order to match the nonlinearities of interest. Since identifying discrete-time Volterra series from training signals can be regarded as solving a set of linear equations, iterative as well as standard least-squares (LS) approaches are appropriate. Iterative approaches like least mean squares (LMS)~\cite[Sec.~3]{zaknich2005principles} or recursive least squares (RLS)~\cite[Sec.~III]{xia2007nonlinear}  are suitable, if channel dynamics call for frequent kernel update. Standard LS approaches like pseudo-inverses~\cite{golub1965calculating} are suitable, if a sufficiently large data set is available and if channel dynamics are nearly static.
They invert matrices, built from transmitted and received training data \cite{ghannouchi2015behavioral}. It has been shown \cite{raich2004orthogonal} that these matrices can be ill-conditioned and therefore become almost singular with growing VNLE complexity.
As a result, the computation of its inverse is prone to large numerical errors. This property calls for a trade-off between matrix conditioning and the number of kernels for maximum performance. In order to avoid the problem of ill-conditioned entirely, abovementioned iterative approaches are applied in this paper. In particular, we use gradient descent and consider two objective functions. First, we follow the standard approach and minimize symbolwise mean square errors (MSE) at the VNLE output. Second, we consider as objective a bitwise soft criterion, which we discuss in detail in Sec.~\ref{sec:bitwise loss tt}.

\subsection{Principles of Deep Neural Network Equalizers} \label{Principles of Deep Neural Network Equalizers}
DNNs computing structures are built from several layers of artificial neurons~\cite[Chap.~5]{bishop2006pattern}.
\begin{figure}
\centering

\begin{tikzpicture}[
	node distance = 3mm and 16mm,
	start chain = going below,
	arro/.style = {-Latex},
	bloque/.style = {text width=6ex, inner sep=2pt, align=right, on chain},
	]
	\foreach \i [count=\j] in {1, 2, s}
	\node[bloque] (in-\j) {$a^{[\ell-1]}_{\i}$};
	\node (out) [circle, draw=orange, minimum size=6mm,right=of $(in-1)!0.5!(in-3)$]  {$\sum$ $\sigma$};
	\foreach \i in {1,...,1}
	\draw[arro] (in-\i) -- (out) node[text width=0.8cm,midway,above=0.1em ] {$w_1$};
	\foreach \i in {2,...,2}
	\draw[arro] (in-\i) -- (out) node[text width=0.8cm,midway,above=-0.2em ] {$w_2$};
	\foreach \i in {3,...,3}
	\draw[arro] (in-\i) -- (out) node[text width=0.8cm,midway,above= -0.35em ] {$w_s$};
	\draw[dotted] (in-2) -- (in-3);
	\coordinate[right=0.3cm of out] (output);
	\draw[arro] (out) -- (output) node[right]   {$\begin{aligned} a^{[\ell]} = \sigma\bigg(\sum_{s=1}^S a^{[\ell-1]}_{s} w_{s} + b \bigg) \end{aligned}$};
	
	\node[above=of in-1.center]     {Input};
	\node[above=of in-1 -| out]     {Projection};
	\node[above right=of in-1 -| output]  {Output};
	\draw (2.2,-1.46) -- (2.2,-0.38);
	\draw [arro] (2.15,-0.1) -- (out); 
	\node[] at (2.15,0.1)     {$b$};
\end{tikzpicture}
\vspace{-0.7cm}

\caption{Single artificial neuron.}
\vspace{-0.5cm}
\label{Single artificial neuron}
\end{figure}
A single artificial neuron is a processing unit with a number of inputs and one output, as shown in Fig.~\ref{Single artificial neuron}. Each input is associated  with a weight. The neuron first computes an activation by summing up the particular weighted inputs and a bias term. Secondly, an activation function~$\sigma(\cdot)$ is applied to obtain the neuron’s output. Interconnecting multiple neurons leads to a DNN. 
With $\boldsymbol{y}= [y(k-M),\dotsc,y(k),\dotsc,y(k+M)]$ and $\Tilde{y}(k)$ representing delayed signal input vector and scalar output, respectively, the DNNE with $L$-layers is given by 
\begin{align}
\boldsymbol{a}^{[0]} &=\boldsymbol{y},
\label{eq.: dnn_line3}\\
\boldsymbol{a}^{[l]} &= \sigma(\boldsymbol{W}^{[l]} \boldsymbol{a}^{[l-1]}+\boldsymbol{b}^{[l]}),\quad l=1,\dotsc,L
\label{eq.: dnn_line7}\\
\Tilde{y}(k) &= \boldsymbol{a}^{[L]}
\label{eq.: dnn_line5}
\end{align}
where $\boldsymbol{a}^{[l]}$ denotes the output vector of the $l$-th layer and $\boldsymbol{W}^{[l]}$ and $\boldsymbol{b}^{[l]}$ the weight matrices and bias vectors, respectively. For the activation function $\sigma$, we use in this work the non-linear function $\sigma(x)=\tanh(x)$ for the hidden layers and the linear function $\sigma(x)=x$ for the output layer (we discuss computationally efficient alternatives to $\tanh$ in Sec.~\ref{sec:dnn complexity}). Eq.~\eqref{eq.: dnn_line3} denotes the input layer and \eqref{eq.: dnn_line7} is executed successively for layers $l=1,2,\dotsc,L$ to obtain output $\Tilde{y}(n)$ in \eqref{eq.: dnn_line5}. While the VNLE memory length of each particular order can be adjusted individually, the DNNE memory length $M$ is a single parameter. 
In the following, \eqref{eq.: dnn_line3},~\eqref{eq.: dnn_line7}, and~\eqref{eq.: dnn_line5} will be referred by
\begin{equation}
\tilde{y}(n) = f_\text{DNNE}(\boldsymbol{y},\boldsymbol{W}^{[1]},\dotsc,\boldsymbol{W}^{[L]},\boldsymbol{b}^{[1]},\dotsc,\boldsymbol{b}^{[L]}).
\label{DNNE-base}
\end{equation} 

Before operations, as in the case of the VNLE, the weights and biases have to be identified. In comparison to the VNLE, only iterative training approaches are appropriate. The iterative training is  based on gradient descent in combination with the backpropagation algorithm~\cite{hecht1992theory}\cite[Sec.~6.5]{goodfellow2016deep}. The weights and biases are updated by shifting previous values towards the gradient descent of the iteratively calculated loss function. To move gently towards the global minimum the adaptive moment estimation (ADAM) \cite{kingma2014adam} is used.

\subsection{Volterra Series and Neural Networks} \label{Volterra Series and Neural Networks}
While VNLEs represent the solutions of nonlinear differential systems based on its Volterra series and hence model nonlinearities with polynomials, DNNEs compute the solutions to a large class of general nonlinear systems on basis of the nonlinear activation functions. %
To depict the relation and the capability of replacing VNLEs by DNNEs, we follow the approach of D. I. Soloway \textit{et al.} \cite{soloway1992neural} and expand the trained DNNE into a Volterra series. This is possible if the activation functions and hence their compositions are infinitely differentiable. The expansion enables a comparison of the linear and nonlinear kernels of both equalizers based on the Volterra series. The dependency between the DNNE parameters and the Volterra kernels is given at a specific input vector $\mathbf{y_0}$ by the partial derivations. For order one and two, the gradient and Hessian matrix are respectively 
\begin{align}
\boldsymbol{h}_1 &= \nabla f_{\text{DNNE}}(\boldsymbol{y}) \bigg\rvert_{\boldsymbol{y} = \boldsymbol{y_0}},
\label{gradient}\\ 
\boldsymbol{h}_2 &= \text{Hess}\left[f_{\text{DNNE}}(\boldsymbol{y}) \right]\bigg\rvert_{\boldsymbol{y} = \boldsymbol{y_0}},
\label{hessian}
\end{align}
\vspace{-0.0cm}
where 
\vspace{-0.0cm}
\begin{align}
\boldsymbol{h}_1 &= [h_1(-M),\dotsc,h_1(M)],\label{eq:kernel order 1}\\
\boldsymbol{h}_2 &= \begin{bmatrix} 
h_p(-M,-M) & \dots & h_p(-M,M) \\
\vdots & \ddots & \vdots \\
0 &   \dots     & h_p(M,M)
\end{bmatrix}.\label{eq:kernel order 2}
\end{align}
The obtained matrix which represents the 2nd order is an upper triangular matrix according to~(\ref{eq.:VNLE}), terms like $y(k)y(k-1)$ and $y(k~-~1)y(k)$ are identical, so that $h_2(0,1)$ and $h_2(1,0)$ can be combined to one unique kernel. Similarly, for order $p$, the kernel is
\vspace{-0.4cm}
\begin{align}
\boldsymbol{h}_p &= \frac{1}{p!} \nabla^p f_{\text{DNNE}}(\boldsymbol{y}) \bigg\rvert_{\boldsymbol{y} = \boldsymbol{y_0}}. \label{gradient_high}
\end{align}
Generalizing \eqref{eq:kernel order 1} and \eqref{eq:kernel order 2}, we can represent the order $p$ kernel by the set
\begin{align}
\boldsymbol{h}_p=\left \{ h_p(s_s,\dotsc,s_p)\colon -M\leq s_1\leq s_2\leq\dotsb\leq s_p\leq M\right \} .\nonumber
\end{align}

\vspace{-0.2cm}
\subsection{MSE for Training Nonlinear Equalizers~\cite{bluemm2019equalizing}}
 \label{SDNNE}
Fig.~\ref{blockdiagram_DNNE_architectures} shows a previously published concept~\cite{bluemm2019equalizing} of replacing a VNLE one-to-one by a DNNE. Symbols $x(i),$ $i = 1 \hdots n$ are transmitted over a noise channel. An observation $y(i)$ is presented to the VNLE or DNNE, which output the equalized signal $\tilde{y}(i)$. 
\tikzset{
	label/.style = {draw, fill=white, rectangle, minimum height=2em, minimum width=2em},
	block/.style = {draw, fill=white, rectangle, minimum height=2em, minimum width=3.5em},
	tmp/.style  = {coordinate}, 
	sum/.style= {draw, fill=white, circle, node distance=1cm},
	input/.style = {coordinate},
	output/.style= {coordinate},
	pinstyle/.style = {pin edge={to-,thin,black}
	}
}
\begin{figure}[b]
	\vspace{-0.4cm}
	\footnotesize
	\centering
	\begin{tikzpicture}[auto, node distance=0.2cm,>=latex']
	\node [label, draw=white] (var) {(a)};
	\node [label, draw=white,right=-0.1cm of var] (input) {$x(i)$};
	\node [block, right=0.2cm of input] (channel) {channel};
	\node [label, draw=white,right=0.2cm of channel] (channeloutput) {$y(i)$};
	\node [block, right=0.2cm of channeloutput] (VNLE) {VNLE};
	\node [label, draw=white,right=0.2cm of VNLE] (VNLEoutput) {$\tilde{y}(i)$};
	\node [block, right=0.2cm of VNLEoutput] (SD) {HD};
	\node [label, draw=white,right=0.2cm of SD] (SDoutput) {${\hat{x}}(i)$};
	\draw [->] (input) -- (channel);
	\draw [->] (channel) -- (channeloutput);
	\draw [->] (channeloutput) -- (VNLE);
	\draw [->] (VNLE) -- (VNLEoutput);
	\draw [->] (VNLEoutput) -- (SD);
	\draw [->] (SD) -- (SDoutput);
	
	\node [label, draw=white,below=0.3cm of var] (var1) {(b)};
	\node [label, draw=white,below=0.3cm of input] (input1) {$x(i)$};
	\node [block, below=0.3cm of channel] (channel1) {channel};
	\node [label, draw=white,below=0.3cm of channeloutput] (channeloutput1) {$y(i)$};
	\node [block, below=0.3cm of VNLE] (VNLE1) {DNNE};
	\node [label, draw=white,below=0.3cm of VNLEoutput] (VNLEoutput1) {$\tilde{y}(i)$};
	\node [block, below=0.3cm of SD] (SD1) {HD};
	\node [label, draw=white,below=0.3cm of SDoutput] (SDoutput1) {${\hat{x}}(i)$};
	\draw [->] (input1) -- (channel1);
	\draw [->] (channel1) -- (channeloutput1);
	\draw [->] (channeloutput1) -- (VNLE1);
	\draw [->] (VNLE1) -- (VNLEoutput1);
	\draw [->] (VNLEoutput1) -- (SD1);
	\draw [->] (SD1) -- (SDoutput1);

	\end{tikzpicture}
	\vspace{-0.0cm}
	\caption{Channel with a (a) VNLE or (b) DNNE accompanied by a Hard decision (HD)} demapper.
	\label{blockdiagram_DNNE_architectures}
\end{figure}
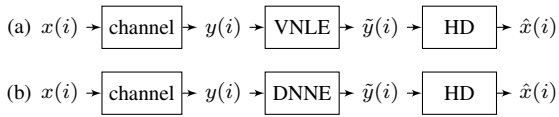
In a simple setting, with memory length equal to one, the observation $y(i)$ is the real valued channel output at the $i$-th channel use. In a more complex setting, with assigned memory taps, $y(i)$ represents several successive channel outputs before and after the $i$-th channel use.

It has been shown that the DNNE trained with respect to the mean-square error (MSE) between transmitted $x(i)$ and received signal $\tilde{y}(i)$ effectively performs symbol-wise hard decisions. While an improvement in hard-decision~(HD) BER in comparison to the VNLE could be achieved the following post-FEC BER of a soft-decision (SD) FEC decoder would be suboptimal.

\vspace{0.4cm}
\subsection{Bitwise Loss Function for Training Soft-Demappers}  \label{sec:bitwise loss tt}
To enable soft-decision FEC decoding, a soft-demapper calculates soft-bits  $\ell_i$, whose sign represent the hard decision, i.e., $0$ if $\ell_i>0$ and $1$ if $\ell_i<0$, and whose absolute values $|\ell_i|$ indicate how confident the soft-demapper is about its decisions, i.e., large $|\ell_i|$ indicate high confidence. By \cite[Eq.~(100)]{bocherer2019probabilistic}, we can estimate from the transmitted bits $b$ and the soft bits $\ell$ the achievable rate per real dimension
\begin{align}
R= m - \min_{s\geq 0}\sum_{i=1}^n\sum_{j=1}^m \log_2\left[1+\exp(-s(1-2b_{ij})\ell_{ij})\right]\label{eq:achievable rate}
\end{align}
where $m$ is the number of bits per real symbol, e.g., $m=3$ for $64$-QAM, where $n$ is the number of transmitted symbols, and where we assumed uniformly distributed channel input symbols. Note that \eqref{eq:achievable rate} is called GMI in \cite[Eq.~(32)]{alvarado2015replacing}.

\begin{figure}[h!]
\footnotesize
	\centering
	\vspace{-0.1cm}	
	\begin{tikzpicture}[auto, node distance=0.3cm,>=latex']
	\node [label, draw=white] (var) {(c)};
	\node [label, draw=white,right=-0.1cm of var] (input) {$x(i)$};
	\node [block, right=0.2cm of input] (channel) {channel};
	\node [label, draw=white,right=0.18cm of channel] (channeloutput) {$y(i)$};
	\node [block, right=0.18cm of channeloutput] (VNLE) {VNLE};
	\node [label, draw=white,right=0.2cm of VNLE] (VNLEoutput) {$\tilde{y}(i)$};
	\node [block, right=0.2cm of VNLEoutput] (SD) {SD};
	\node [label, draw=white,right=0.2cm of SD] (SDoutput) {$\ell_1(i),\dotsc,\ell_m(i)$};
	\draw [->] (input) -- (channel);
	\draw [->] (channel) -- (channeloutput);
	\draw [->] (channeloutput) -- (VNLE);
	\draw [->] (VNLE) -- (VNLEoutput);
	\draw [->] (VNLEoutput) -- (SD);
	\draw [->] (SD) -- (SDoutput);
	\end{tikzpicture}
\end{figure}
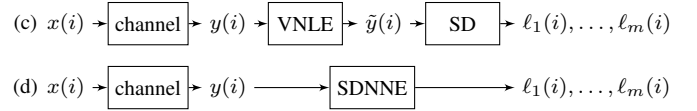
\begin{figure}[h!]
\footnotesize
	\centering
	\vspace{-0.5cm}	
	\begin{tikzpicture}[auto, node distance=0.3cm,>=latex']
	\node [label, draw=white] (var) {(d)};
	\node [label, draw=white,right=-0.1cm of var] (input) {$x(i)$};
	\node [block, right=0.2cm of input] (channel) {channel};
	\node [label, draw=white,right=0.18cm of channel] (channeloutput) {$y(i)$};
	\node [block, right=1cm of channeloutput] (SDNNE) {SDNNE};
	\node [label, draw=white,right=1.325cm of SDNNE] (SDNNEoutput) {$\ell_1(i),\dotsc,\ell_m(i)$};
	\draw [->] (input) -- (channel);
	\draw [->] (channel) -- (channeloutput);
	\draw [->] (channeloutput) -- (SDNNE);
	\draw [->] (SDNNE) -- (SDNNEoutput);
	\end{tikzpicture}
	\vspace{-0.3cm}
	\caption{Channel with a (c) VNLE accompanied by a Soft-decision (SD) demapper or a (d) Soft Deep Neural Network Equalizer (SDNNE).}
	\label{channel_NN_demapper}
	\vspace{-0.2cm}
\end{figure}
We now use the DNNE as a soft-demapper. To this end, we define $m$ output units with linear activation functions, to allow for negative and positive values. To maximize the achievable rate \eqref{eq:achievable rate}, we use as loss function the bitwise equivocation
\begin{align}
\mathcal{L}(b, \ell)=\log_2[1+\exp(-(1-2b)\ell)]\label{loss function}
\end{align}
where $b$ is the transmitted bit and where $\ell$ is the corresponding SDNNE output. In the appendix, we show the equivalence of \eqref{loss function} to the binary cross-entropy used in classic machine learning for binary classification and we also show that this loss function is optimal, in the sense that a trained SDNNE maximizes the achievable rate \eqref{eq:achievable rate}. Note that \eqref{loss function} does not include the minimization over $s$ that we have in \eqref{eq:achievable rate}. This allows us to check if training has been successful. For the soft bits output by the trained SDNNE, the minimizing $s$ in \eqref{eq:achievable rate} should be equal to $1$. Otherwise, further training is required.

The VNLE followed by a soft-demapper realizes the same functionality as the SDNNE. To optimize the VNLE with respect to the bitwise equivocation \eqref{loss function}, we realize the soft-demapper by the max-log approximation, which incurs virtually no loss in the SNR ranges considered in this work. The MLA effectively uses piecewise linear approximations, whose slopes form the MLA parameters. We then optimize the MLA and the VNLE jointly by gradient descent,  minimizing the bitwise equivocation \eqref{loss function}. Note that by doing so, both the SDNNE and the VNLE followed by the soft-demapper are optimized with respect to the same bitwise soft criterion.

\subsection{Architecture Design of Deep Neural Network Equalizers} \label{Architecture Design of Deep Neural Network Equalizer}
While the design of the input and output layer depends on the input and output dimension of the desired function, the design options of the hidden layers are numerous and interrelated. However, in the last decade, the amount of publications, which investigated neural networks from a theoretical perspective, have increased.

Mont{\'u}far \textit{et al.} (2014) \cite{montufar2014number} demonstrated that deep networks are more expressive and require far fewer neurons to represent a desired function. He showed that the number of decision regions of deep models grow exponentially in the depth and polynomially in the number of neurons in the hidden layers, which is much faster than that of shallow models which grow only polynomially instead of exponentially in the number of hidden units~\cite{barron1993universal}.
Consequently, deeper models need exponentially less parameters to reach the desired representation capacity. Choosing a deeper model and hence a cascade structure matches also better with the intertwined memory effects and nonlinearities in the physical channel. 
Deeper layers can reuse constructed conclusions from the lower layers in order to build gradually more complex functions \cite{pascanu2013number}.

Furthermore, to assess the number of neurons in the hidden layers, Mont{\'u}far \textit{et al.} (2017) \cite{montufar2017notes} introduced an upper bound on the number of activation patterns. In a neural network the number of activation patterns correspond to the number of possible distinct input space regions that the neural network can distinguish.
For fixed parameters, a network with $n_0$ inputs and $n_1,\dotsc,n_L$ neurons in the hidden layers realizes at most
\begin{equation}
\prod_{l=1}^{L} \sum_{j=0}^{m_l-1} {n_l \choose j}, \qquad m_{l-1} = \min \{n_0,\dotsc,n_{l-1} \} \label{upper_bound}
\end{equation}
activation patterns as one traverses the input space. The bound provides an indication of the representation capacity and hence of the capacity of a SDNNE architecture. It helps to assess the number of neurons in the hidden layers for a given equalization scenario.

\tikzset{%
	every neuron/.style={
		circle,
		draw=orange,
		minimum size=0.6cm,
	},
	neuron missing/.style={
		draw=none, 
		scale=1.5,
		text height=0.3cm,
		execute at begin node=\color{black}$\vdots$
	},
}
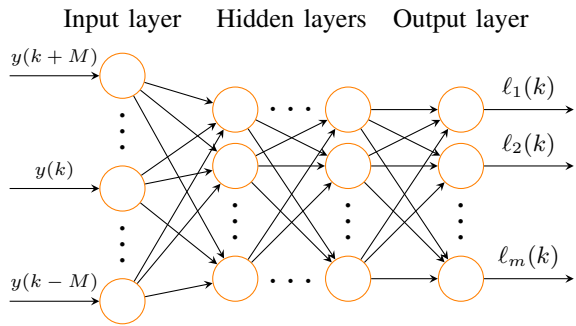
\begin{figure}[h!]
	\begin{center}
		\begin{tikzpicture}[x=1.5cm, y=1.5cm, >=stealth]
		
		\foreach \m/\l [count=\y] in {1,missing,3,missing,5}
		\node [every neuron/.try, neuron \m/.try] (input-\m) at (0,2-\y*0.5) {};
		
		\foreach \m [count=\y] in {1,2,missing,4}
		\node [every neuron/.try, neuron \m/.try ] (hidden1-\m) at (1,1.7-\y*0.5) {};
		
		\foreach \m [count=\y] in {1,2,missing,4}
		\node [every neuron/.try, neuron \m/.try ] (hidden2-\m) at (2,1.7-\y*0.5) {};
		
		\foreach \m [count=\y] in {1,2,missing,4}
		\node [every neuron/.try, neuron \m/.try ] (output-\m) at (3,1.7-\y*0.5) {};
		
		\foreach \l [count=\i] in {1}
		\draw [<-] (input-\i) -- ++(-1,0)
		node [above, midway] {\scriptsize $y(k+M)$};
		
		\foreach \l [count=\i] in {3}
		\draw [<-] (input-3) -- ++(-1,0)
		node [above, midway] {\scriptsize $y(k)$};
		
		\draw [<-] (input-5) -- ++(-1,0)
		node [above, midway] {\scriptsize $y(k-M)$};
		
		\foreach \l [count=\i] in {1,2}
		\draw [->] (output-\i) -- ++(1,0)
		node [above, midway] {\small $\ell_\i(k)$};

		\draw [->] (output-4) -- ++(1,0)
		node [above, midway] {\small $\ell_{m}(k)$};

		\foreach \i in {1,3,5}
		\foreach \j in {1,2,4}
		\draw [->] (input-\i) -- (hidden1-\j);
		
		\foreach \i in {1,2,4}
		\foreach \j in {2}
		\draw [->] (hidden1-\i) -- (hidden2-\j);
		
		\draw [->] (hidden1-2) -- (hidden2-1); 
		\draw [->] (hidden1-4) -- (hidden2-1); 
		
		\draw [->] (hidden1-1) -- (hidden2-4); 
		\draw [->] (hidden1-2) -- (hidden2-4); 
		
		\foreach \i in {1,2,4}
		\foreach \j in {1,2,4}
		\draw [->] (hidden2-\i) -- (output-\j);
		
		\node [align=center, above,scale=1.5] at (1.52,1.06) {$\dots$};
		\node [align=center, above,scale=1.5] at (1.52,-0.46) {$\dots$};

		\foreach \l [count=\x from 0] in {Input layer, Hidden layers, Output layer}
		\node [align=center, above] at (\x*1.5,1.8) {\l};
		
		\end{tikzpicture}
		\caption{Deep Neural Network Equalizer for separate in phase~(I) and quadrature~(Q) processing (four times required in coherent, dual polarization systems). The architecture deploys a dual-side symmetric memory structure where $M$ denotes the number of single side memory taps and $m$ the number of bits per real symbol.}\label{NN_struc}
	\end{center}
\end{figure}

\begin{figure}[t!]
	\begin{center}
		\includegraphics[width=3.5in]{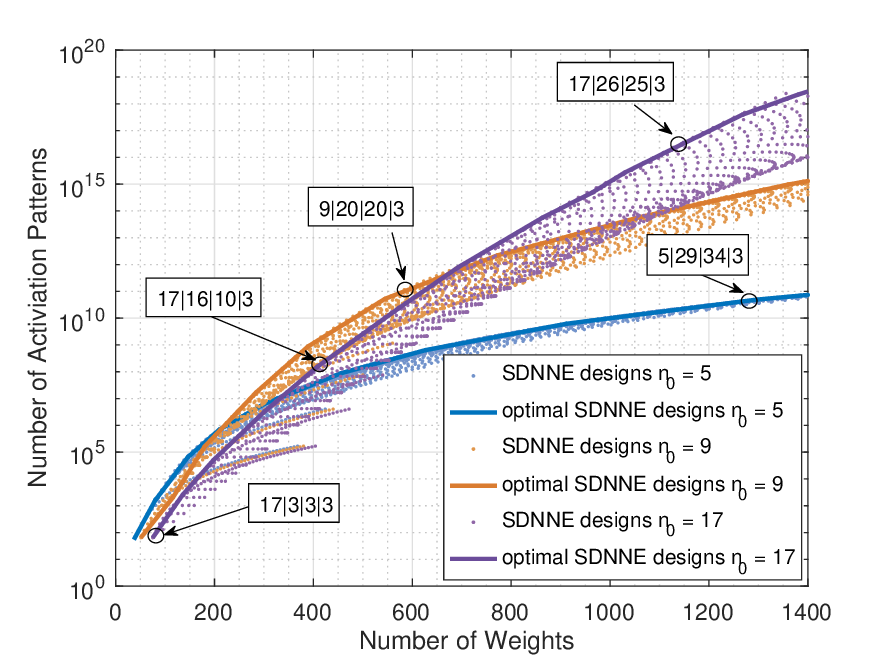}\\
		\caption{Number of activation patterns of a SDNNEs with different memory sizes and two hidden layers with various neurons. Some examplary optimal designs are labeled with their structures, where the numbers denote the number of neurons. A detailed definition of the structures can be found in Sec.~\ref{Performance Evaluation}}\label{Number_of_activation_patterns}
		\vspace{-0.55cm}
	\end{center}
\end{figure}
Fig.~\ref{NN_struc} illustrates the structure of an SDNNE with multiple hidden layers, which is applied separately to the in phase (I) and quadrature (Q) components of the two polarizations.
Equal to the VNLE design, the memory effects of channel and components are considered by adding time delayed versions of the input signal. For the hidden neurons $\tanh$ has been chosen as activation function, due to its differentiable property described in Sec.~\ref{Volterra Series and Neural Networks}. In comparison to the commonly used sigmoid function, the linear approximation of a $\tanh$ function is the straight line passing through the neighbourhood of zero~\cite{soloway1992neural}. This property facilitates the Volterra series expansion of the SDNNE, which we use in Sec.~\ref{subsec:VNLE and SDNNE kernels} for comparison with VNLE kernels.

Fig.~\ref{Number_of_activation_patterns} shows the number of activation patterns related to the number of weights of different SDNNE architectures with various input neurons, i.e., memory sizes. Each small dot denotes a particular SDNNE architectural design.  
It can be observed that for each memory size an architectural trade-off between complexity and representation capacity exists. The optimal architectures are obviously located at the corresponding envelops and are therefore the preferred choices. Some example optimal designs are labeled with their structures, which we define in Sec.~\ref{Performance Evaluation}.

\tikzset{
	triangle/.style = {draw, fill=white, regular polygon, regular polygon sides=3},
	node rotated/.style = {rotate=180},
	border rotated/.style = {shape border rotate=180},
	triangleEDFA/.style = {draw, fill=white, regular polygon, regular polygon sides=3,minimum size=1.2cm},
	node1 rotated/.style = {rotate=270},
	border1 rotated/.style = {shape border rotate=270},
	label/.style = {draw, fill=white, triangle, minimum height=2em, minimum width=5em},
	VOA/.style = {draw, fill=white, circle, minimum size=1.5em},
	blockDSPTX/.style = {draw, fill=white, rectangle, minimum height=1.2em, minimum width=7em},
	blockDSPRX/.style = {draw, fill=white, rectangle, minimum height=1em, minimum width=12em},
	VNLE/.style = {draw, fill=white, rectangle, minimum height=1.2em, minimum width=5em},
	blockOptical/.style = {draw, fill=white, rectangle, minimum height=2.8em, minimum width=4em},
	PD/.style = {draw, fill=white, rectangle, minimum height=1.2em, minimum width=1.2em},
	mul/.style = {draw, fill=white, circle, node distance=0.2cm},
	tmp/.style  = {coordinate}, 
	sum/.style= {draw, fill=white, circle, node distance=1cm},
	input/.style = {coordinate},
	output/.style= {coordinate},
	pinstyle/.style = {pin edge={to-,thin,black}
		
	}
}
\begin{figure*}
	\begin{tikzpicture}[auto, node distance=0.2cm,>=latex']
	
	\node[inner sep=0pt] (foto4) at (15.5,-1.6)
	{\includegraphics[width=.13\textwidth]{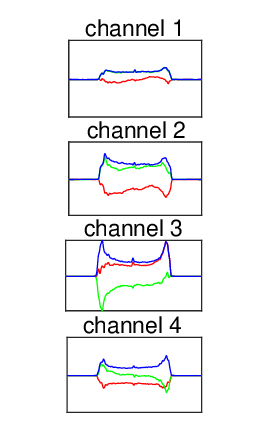}};
	\draw[-,decorate,decoration={brace,mirror}] (14.8,0.1) -- node[midway, above,yshift=-6pt,xshift=-18pt] {\footnotesize} (14.8,-3.4);

	\node [blockDSPTX] (Mapping) {\footnotesize Mapping};
	\node [blockDSPTX, below=0.05cm of Mapping] (Pilot Insertion) {\footnotesize Pilot Insertion};
	\node [blockDSPTX, below=0.05cm of Pilot Insertion] (Pulse Shaping) {\footnotesize Pulse Shaping};
	\node [blockDSPTX, below=0.05cm of Pulse Shaping] (Preamphasis) {\footnotesize Preemphasis};
	\node [blockDSPTX, below=1.0cm of Preamphasis] (DAC) {\footnotesize 100 GSa/s DACs};
	
	\node [blockDSPTX, below=0.05cm of DAC] (DA) {\footnotesize Driver Amplifiers};
	
	\node [blockOptical,minimum width=2em] at (0.8,-5.35) (DP-IQM) {\footnotesize DP-IQM};
	\node [blockOptical,align=center,minimum width=1.5em] at (-0.7,-5.35)  (TECL) {\footnotesize Tunable\\ \footnotesize ECL};
	\node [triangleEDFA, border1 rotated,right=0.3cm of DP-IQM] (EDFA1) {};
	\node [triangleEDFA, border1 rotated,above=0.55cm of EDFA1] (EDFA3) {};
	\node [VOA,right=0.25cm of EDFA1] (VOA) {};
	\node [VOA,above=0.7cm of VOA] (VOA3) {};
	\node [blockOptical, right=0.25cm of VOA,align=center] (Coupler) {\footnotesize Coupler};
	\node [blockOptical, right=0.5cm of Coupler,align=center] (Mux) {\footnotesize 100-GHz\\ \footnotesize Mux};
	\node [triangleEDFA, border1 rotated,right=0.4cm of Mux] (EDFA2) {};
	\node [VOA,right=0.25cm of EDFA2] (VOA1) {};
	\node [blockOptical, right=0.25cm of VOA1,align=center] (DeMux) {\footnotesize 100-GHz\\ \footnotesize DeMux};
	\node [blockOptical,align=center,minimum height=3em,minimum width=3em] at (9.3,-4) (OSA) {\footnotesize OSA};	
	\node [blockOptical,align=center,minimum height=1em,minimum width=6.3em] at (12.5,-5.6) (Hybrid) {\footnotesize $90^{\circ}$Hybrid};
	\node [blockOptical, right=3.5cm of DeMux,align=center] (TECL2) {\footnotesize Tunable\\ \footnotesize ECL};
	\node [blockOptical, above=0.05cm of Hybrid,align=center,minimum height=1em,minimum width=6.3em] (PD1) {\footnotesize Photodiodes};
	\node [blockOptical, above=0.05cm of PD1,align=center,minimum height=1em,minimum width=6.3em] (ADC) {\footnotesize 256GSa/s Scope};
	\node [blockDSPRX, above=0.19cm of ADC,align=center] (CDC) {\footnotesize CD Compensation};
	\node [blockDSPRX, above=0.04cm of CDC,align=center] (CFO) {\footnotesize Coarse CFO Compensation};
	\node [blockDSPRX, above=0.04cm of CFO,align=center] (Framing) {\footnotesize Framing};
	\node [blockDSPRX, above=0.04cm of Framing,align=center] (Fine CFO) {\footnotesize Fine CFO Compensation};
	\node [blockDSPRX, above=0.04cm of Fine CFO,align=center] (MIMO) {\footnotesize $2 \times 2$ MIMO};
	\node [blockDSPRX, above=0.04cm of  MIMO,align=center] (TCR) {\footnotesize Timing \& Carrier Recovery};
	\node [VNLE, right=9.16cm of Mapping,align=center] (SD) {\footnotesize SD};
	\node [VNLE, below=0.05cm of SD,align=center] (VNLE) {\footnotesize VNLE};
	\node [VNLE, right=0.7cm of VNLE,align=center] (SDNNE) {\footnotesize SDNNE};
	
	\node[text width=1cm] at (10.5,-0.25) {\footnotesize (a)};
	\node[text width=1cm] at (12.9,-0.5) {\footnotesize (b)};
	
	\node[inner sep=0pt] (foto3) at (3.1,-1)
	{\includegraphics[width=.20\textwidth]{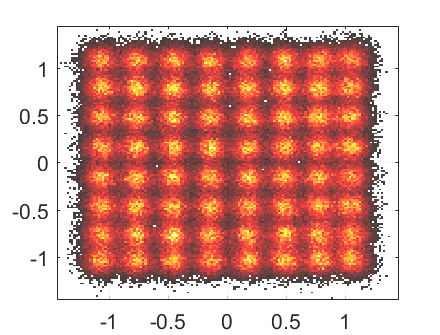}};
	\draw [<-] [dashed] (10.5,-0.82) -- (10.2,-0.82)-- (10.2,-2.5)-- (3,-2.5)-- (3,-2.2);
	
	\node[inner sep=0pt] (foto4) at (7.5,-1)
	{\includegraphics[width=.27\textwidth]{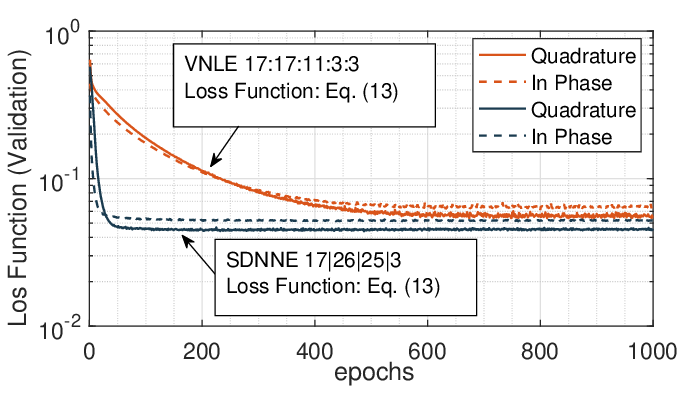}};	
	
	\node[inner sep=0pt] (foto5) at (6.5,-3.7)
	{\includegraphics[width=.23\textwidth]{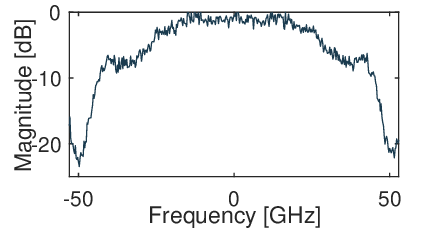}};	
	\draw [<-] [dashed] (12,-4.25) -- (10,-4.25) -- (10,-3) -- (8.2,-3);

	\node[text width=1cm] at (2.25,-4.07) {\footnotesize EDFA};
	\node[text width=1cm] at (2.25,-5.34) {\footnotesize EDFA};
	\node[text width=1cm] at (7.95,-5.34) {\footnotesize EDFA};	
	
	\draw [->] (2.9,-4.4) -- (3.5,-3.8);
	\draw [->] (2.9,-5.65) -- (3.5,-5.05);
	\draw [->] (8.6,-5.65) -- (9.2,-5.05);
	
	\draw [->,thick] (Preamphasis) -- (DAC);
	\draw [->,thick] (DA) -- (0,-4.2) -- (0.8,-4.2)-- (DP-IQM);
	\draw [dashed] (1.5,-3.5) -- (4,-3.5) -- (4,-4.7) -- (1.5,-4.7) -- (1.5,-3.5);\
	\node[text width=4cm] at (3.5,-3.35) {\footnotesize ASE Noise Loading};

	\draw [->,thick] (TECL) -- (DP-IQM);
	\draw [->,thick] (DP-IQM) -- (EDFA1);
	\draw [->,thick] (EDFA1) -- (VOA);
	\draw [->,thick] (VOA) -- (Coupler);
	\draw [->,thick] (Coupler) -- (Mux);
	\draw [->,thick] (Mux) -- (EDFA2);
	\draw [->,thick] (EDFA2) -- (VOA1);
	\draw [->,thick] (VOA1) -- (DeMux);
	\draw [->,thick] (OSA) -- (EDFA2);

	\draw [->,thick] (EDFA3) -- (VOA3);
	\draw [->,thick] (VOA3) -| (Coupler);
	
	\draw [->,thick] (DeMux) -| (11.0,-5.6) -- (Hybrid);
	\draw [->,thick] (TECL2) -| (14,-5.6) -- (Hybrid);
	
	\draw [->,thick] (ADC) -- (CDC);
	
	\draw [<-,thick] (SDNNE) -- ++(0,-0.45cm);
	\draw [<-,thick] (VNLE) -- ++(0,-0.45cm);

	\node[text width=1cm] at (6.4,-1.4) {\footnotesize (b)};
	\node[text width=1cm] at (7.35,-0.8) {\footnotesize (a)};

	\end{tikzpicture}

	\caption{Back-to-Back offline measurement setup including Tx and Rx DSP architectures with two options  a) 4x Volterra Equalizers (VNLE) and b) 4x Soft Deep Neural Network Equalizers for separate I and Q processing as well as for each polarization. In addition, the received constellation at 35.2dB OSNR, the learning process, the received spectrum as well as the complex taps of the frequency domain $2\times2$ MIMO equalizer are shown. The four subplots illustrate each polarization and the corresponding mixtures, while green denotes the imaginary, red the real and blue the absolute value of the complex taps.}  
	\label{fig:setup}
	\vspace{-0.2cm}
\end{figure*}
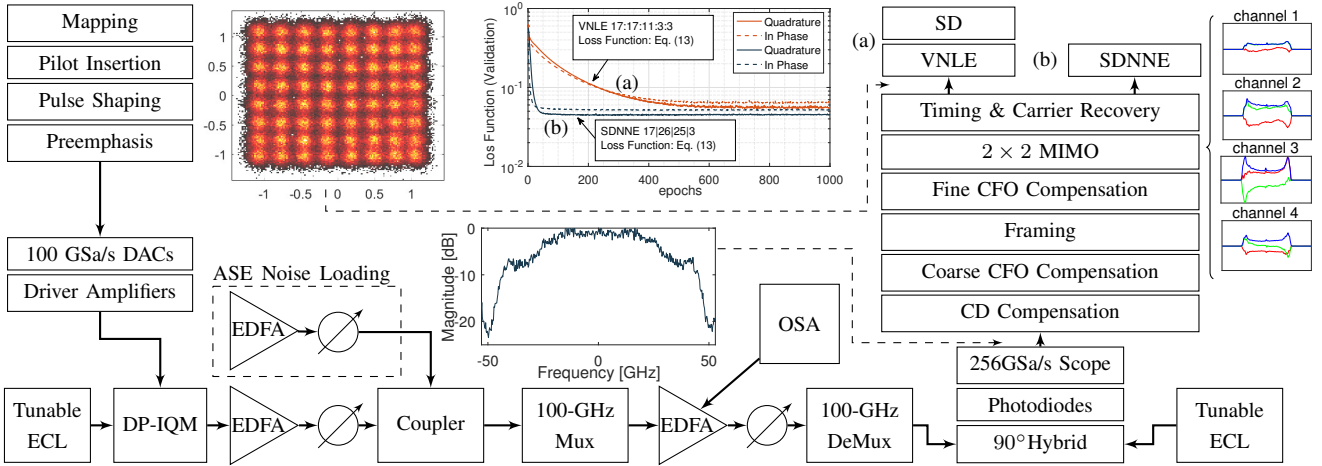

\section{Experimental Comparison of \\ Equalization Performance}
This section outlines the measurement setup, as used to evaluate the nonlinear compensation performance of the proposed architectures on basis of identical offline data. The quality of the obtained soft bits is evaluated with the achievable rate~\eqref{eq:achievable rate}.

\subsection{Experimental Setup}
A coherent single carrier transmission system over a single mode fiber (SMF) is employed to experimentally evaluate the performance of the nonlinear equalizers (NLEs). A schematic of the experimental setup including the offline DSP stack is shown in Fig.~\ref{fig:setup}. 
The setup was optimized for maximum performance without NLEs and the measurements were performed back-to-back with ASE noise loading, in order to compare the achievable rate~\eqref{eq:achievable rate} at varying OSNR values. The signal consists of a 92GBd DP-64QAM with gross data rate of 1104Gb/s. With $1 \%$ for training overhead and assuming an FEC overhead of $15\%$ (e.g., ``oFEC'', \cite[Table~9.1]{jia2020coherent},\cite{openroadm}) and $20\%$ (e.g., LDPC, \cite[Sec.~3]{buchali2014implementation}), the net bit rate is 950Gb/s and 912Gb/s, respectively.
 
At the transmitter, the DSP inserts a constant amplitude zero autocorrelation (CAZAC) training sequence, which is used for framing, carrier frequency offset estimation, residual chromatic dispersion compensation and polarization-decoupling frequency domain $2\times2$ MIMO equalization~\cite{pittala2014training}. The MIMO equalizer considers 51 taps for complex channel estimation and operates with 2 samples per symbol (sps). To compensate for transmitter impairments, a static linear digital preemphasis is applied after pulse-shaping. 
The signal output powers of the four 100GSa/s Micram digital-analog converters (DACs) with 40GHz 3dB-bandwidth and 6-bits nominal resolution are set to -6dBm. Subsequently, the RF signals are amplified by four SHF S804A amplifiers with 22dB gain and 60GHz 3dB-bandwidth. The amplifiers slightly operate in a nonlinear region, which in turn results in nonlinear intermodulation distortions. Their nonlinear effects are mixed with potential nonlinear distortions from a LiNbO3 DP-I/Q Modulator (Fujitsu-FTM7992HM-32GHz) with a drive voltage of $\leq$4.2Vpp.

In the optical domain, two tunable 1kHz external cavity lasers (ECLs) are used at the transmitter and LO, respectively. The optically modulated signal is amplified and then combined with the amplified spontaneous emission (ASE) noise generated by an EDFA. The 100GHz Mux and Demux are considered for future wavelength-division multiplexing~(WDM). The receiver consists of an optical $90^\circ$-hybrid and four 70GHz balanced photodiodes. The photodiodes operate in the linear regime with 0dBm optical input power. The electrical signals are digitized using a Keysight Infiniium real-time oscilloscope including four 10-bits analog-digital converters operating at 256GSa/s with 110GHz 3dB-bandwidth.

In order to compensate transmitter nonlinearities, ISI as well as memory effects, the receiver DSP stack includes next to the classical coherent signal recovery blocks the stacked combination of the VNLEs plus MLA soft-decision demappers and the proposed SDNNEs. Both schemes operate independently on the real dimension with 1 sps. For fair performance comparisons, the different NLE types operate on identical power normalized data. Regardless of the type of NLE in use, training on the particular nonlinearities is essential before deployment. CAZAC sequences do not capture nonlinearities very well. Hence, NLE training is done instead upon $50\%$ of the payload of the first received frame per OSNR value, which consists of 66444 symbols. In order to prevent overfitting, the performance is repetitively validated during the training phase on the remaining $50\%$ of the payload. Once trained, the performance is evaluated on the second half of six new captured frames.

\subsection{Performance Evaluation} \label{Performance Evaluation}
\begin{figure}[t!]
	\begin{center}
		\includegraphics[trim = 0.6cm 0 0 0.0cm,width=3.5in]{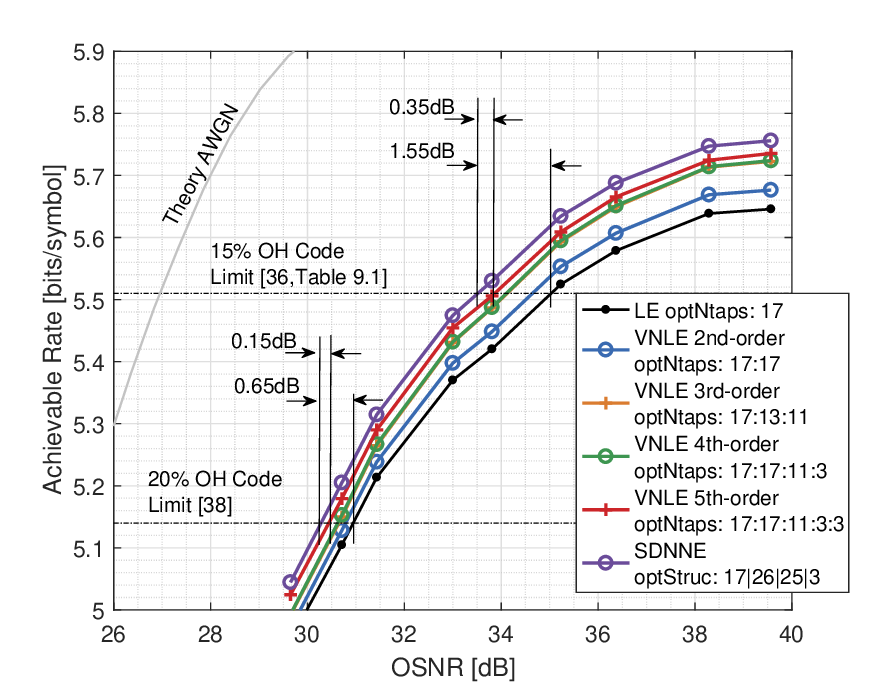}\\		
		\caption{92-Gbaud 64QAM performance in terms of achievable rate versus OSNR for the optical back-to-back system by applying optimized linear as well as nonlinear equalizer.}\label{MI_versus_OSNR}
		\vspace{-0.65cm}
	\end{center}
\end{figure}

\begin{figure}[t!]
	\begin{center}
		\includegraphics[trim = 0.6cm 0 0 0.0cm,width=3.5in]{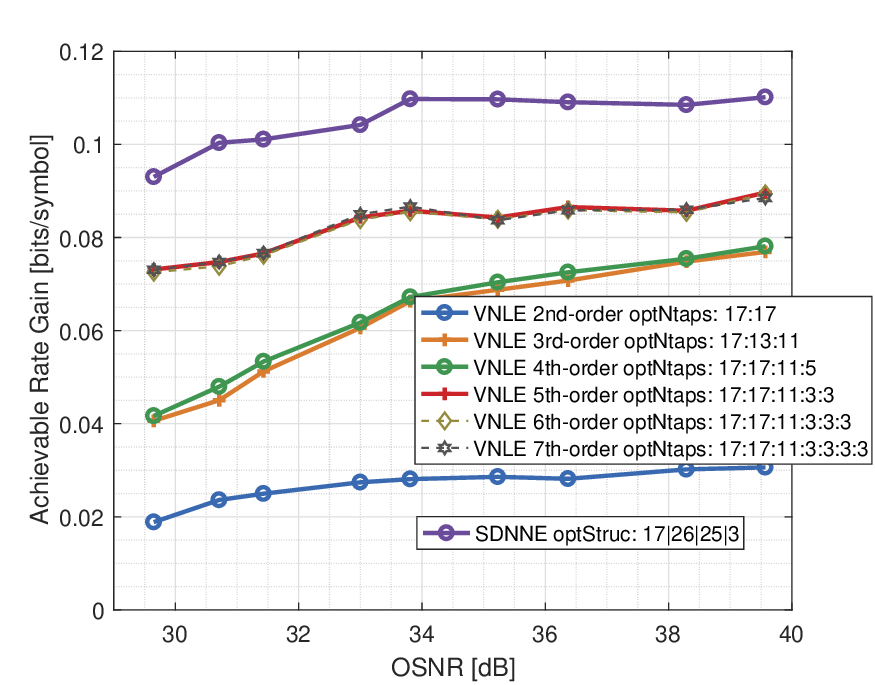}\\
		\caption{Nonlinear compensation gains in bits/symbol related to OSNR for a 92-Gbaud 64QAM optical BtB system. Blue, yellow, green,~red,~gold,~gray represent the VNLE with different orders while the purple represents~the~SDNNE.}\label{GMI_gain}
	\end{center}
	\vspace{-0.8cm}
\end{figure}

Fig. \ref{MI_versus_OSNR} depicts the performance in terms of achievable rate~\eqref{eq:achievable rate} related to OSNR for the optical BtB 92GBd 64QAM measurements by applying a linear equalizer (LE), a VNLE up to 5th order kernels and a SDNNE. Kernel orders greater than 5th, e.g. 6th or 7th, do not yield any additional gain, see Fig.~\ref{GMI_gain}, for readability, we therefore omit these results in Fig.~\ref{MI_versus_OSNR}. The black dot-dashed lines denote the FEC limits for 15\% and 20\% FEC overhead, respectively. All three architectures are optimized regarding achievable rate~\eqref{eq:achievable rate} performance and deploy a dual-side symmetric memory structure, i.e. previously as well as future symbols are taken symmetrically into account. The VNLE design, e.g., $\text{17:17:11:}\dotsc$ stands for 17 linear memory taps (8 previous symbols + the current symbol + 8 future symbols) followed by 17 memory taps for the second order, followed by 11 memory taps for the third order and so on. The SDNNE design, e.g., $17|26|25|3$, stands for 17 input neurons (again 8 previous symbols + the current symbol + 8 future symbols) followed by two hidden layers with 26 and 25 neurons, feeding into 3 output nodes. 

In addition to Fig. \ref{MI_versus_OSNR}, Fig. \ref{GMI_gain} plots the corresponding performance improvement in relation to the LE at various OSNR values. It can be observed that 2nd, 3rd as well as 5th order kernels of the VNLE introduce the major benefit, while the 4th order does not yield significant additional gain. This indicates that odd harmonics dominate. The VNLE of 5th order improves the linear equalized baseline curve at higher OSNR areas up to $\sim 0.09$ bits/symbol. In lower OSNR ranges where ASE noise is the dominant distortion, the gain decreases slightly to $\sim 0.07$ bits/symbol. This behavior applies for all NLE architectures. 
Optimizing the VNLE and the MLA soft demapper jointly with respect to the bitwise equivocation \eqref{loss function} yields an achievable rate improvement of around $0.002$~bits/QAM symbol, compared to the MSE-optimized VNLE. This improvement confirms that the bitwise equivocation is the appropriate objective for maximizing the achievable rate. However, the improvement over the MSE criterion is negligibly small for the scenario considered here.

For the SDNNE architecture different numbers of hidden layers and corresponding numbers of neurons per layer can be examined, in order to optimize the performance. An assessment of the architecture for $17$ memory taps is given by equation~(\ref{upper_bound}) and shown in Fig.~\ref{Number_of_activation_patterns}. To release sufficient representation capacity for the equalization problem, in a first step the more complex design $17|26|25|3$ is chosen. In a second stage, the complexity will be slightly reduced to determine the upper bound of the architecture complexity. It can be observed that the appropriate SDNNE architecture outperforms the 5th order VNLE and improves the linear equalized baseline curve up to $\sim 0.11$ bits/symbol and $\sim 0.09$ bits/symbol at lower OSNR values, respectively.

\subsection{Kernels of Volterra Nonlinear Equalizer and Soft Deep Neural Network Equalizer}
\label{subsec:VNLE and SDNNE kernels}

\begin{figure}[b!]
	\begin{center}
		\vspace{-0.3cm}
		\includegraphics[trim = 0.6cm 0 0 0.7cm,width=3.7in]{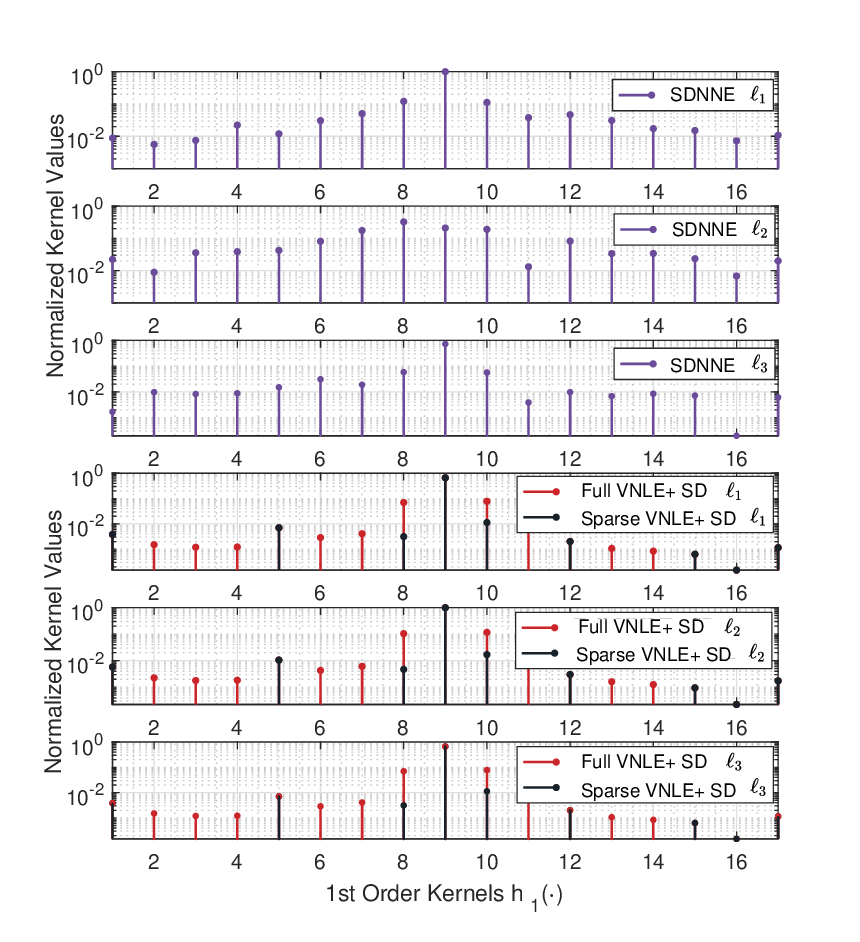}\\
		\vspace{-0.3cm}
		\caption{The linear kernels, or rather the linear finite impulse responses are extracted from the trained SDNNE by Taylor expansion and compared with the trained Volterra kernels. The purple stems represent the bit-level SDNNE linear finite impulse response, while the red and black represent the full and sparse VNLE linear finite impulse response, respectively.}\label{DNN_17_16_10_3_anlyse_1st}
		\vspace{-0.25cm}
	\end{center}
\end{figure}
\begin{figure}[t!]
	\begin{center}
		\includegraphics[trim = 0.6cm 0 0 0.7cm,width=3.7in]{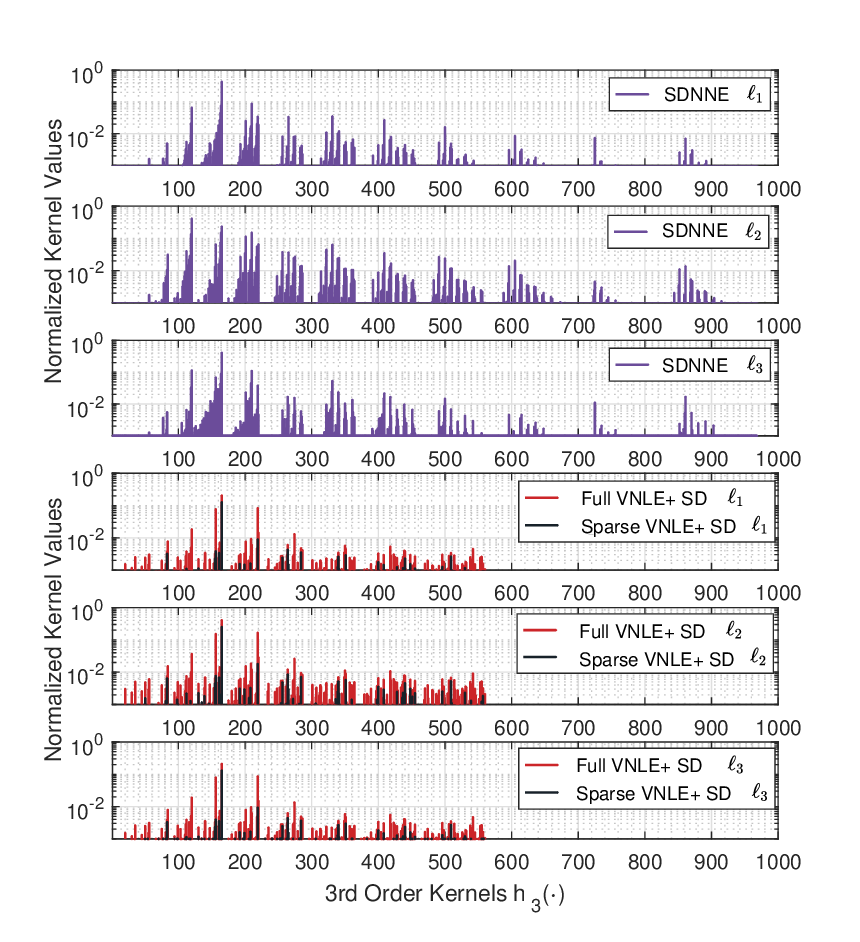}\\
		\vspace{-0.3cm}
		\caption{Comparison of the unrolled extracted third order nonlinear kernels, or rather the third order finite impulse response of the trained SDNNE with the unrolled third order nonlinear kernels of the trained VNLE. }\label{DNN_17_16_10_3_anlyse_3st}
		\vspace{-0.6cm}
	\end{center}
\end{figure}

The Volterra kernels provide a useful tool for analyzing the channel behavior. The first term in expression~(\ref{eq.:VNLE}) represents the common finite linear impulse response of the system, while higher order Volterra kernels represent higher order impulse responses and hence describe the nonlinear dynamic behavior. As described in Sec.~\ref{Volterra Series and Neural Networks}, the trained SDNNE can be expanded into a Volterra series and hence into comparable information. The kernels are expressed in terms of the parameters of the trained SDNNE by using expressions~(\ref{gradient}), ~(\ref{hessian}) and ~(\ref{gradient_high}) at $\mathbf{y_0} = \mathbf{0}$. In comparison to the customary considered symbol based VNLE kernels, the proposed SDNNE provides individual kernels corresponding to the soft outputs. To overcome this representation issue, the VNLE kernels are extended with the particular weights of the max-log soft demapper in order to likewise represent the individual VNLE kernels for the soft bits. Fig.~\ref{DNN_17_16_10_3_anlyse_1st} depicts the in phase linear kernels from the VNLE of polarization~$x$ as well as the corresponding extracted linear kernels of the SDNNE.
In addition, the corresponding kernels of the sparse VNLE are depicted, which we introduce in Sec.~\ref{Complexity Evaluation}.

The linear kernel values from SDNNE $\ell_1$, SDNNE $\ell_3$, VNLE $\ell_1$ and VNLE $\ell_3$ are very similar. Interestingly, the centered bit $\ell_2$, which distinguishes between the inner and outer circles of the constellation points, exhibits higher distortions. 
While the SDNNE is able to adjust the kernels for each soft output individually by adapting the weights of the last layer separately, the VNLE does not exhibits that flexibility. The VNLE kernel values for each soft output only differ in the amplitude.

An additional advantage of an SDNNE is shown in Fig.~\ref{DNN_17_16_10_3_anlyse_3st}. In comparison to the VNLE, where the number of high order taps is limited due to complexity, the SDNNE comprises all possible combinations of the input signal up to the input memory depth. %

\section{Comparison of Complexity} \label{Comparison of Complexity}
As part of the DSP of the optical transceiver, the NLE is only deployable, if it can be implemented efficiently enough on an ASIC. This section compares some complexity aspects of VNLEs versus SDNNEs regarding the main ASIC resources, memory and logic cells on floating point level. Logic size can be well estimated in numbers of multipliers, as they are by far the most expensive logic blocks of VNLEs and SDNNEs. 

In general, the complexity analysis of trainable equalizers includes two aspects, the training and the real-time execution aspect. However, in the case of optical communication, we can assume only insigniﬁcant changes of component nonlinearities over time or slow processes like aging or temperature shifts. Thus, for a VNLE and an SDNNE the extraction of training data can be done offline, once after production, or on an (embedded) microcontroller. Therefore, what matters is the real-time complexity, i.e. the number of calculations that are done on each sample when executing the particular NLE.

\subsection{Complexity of Volterra-based Equalizers and MLA}
The real-time complexity of a VNLE is defined by the number of kernels, see Section~\ref{pV}. Their number is directly connected to the required number of hardware multipliers, namely
\begin{equation}
\text{mul}_\text{VNLE} = \underbrace{N_1 + \sum_{i = 2}^{P} N_i}_{\text{Kernels}} + \underbrace{2M_i+1}_{\text{Feature Matrix\\~\cite[Sec.~IV-B]{tarver2019design}}}.
\label{mul_vnle}
\end{equation}
Eq.~\eqref{mul_vnle} is derived from the structure presented in~\cite[Sec.~IV-B]{tarver2019design}. It considers terms that can be obtained by delaying other terms as well as the reuse of products of order $k$ to compute products of order $k+1$.
Moreover, the accompanied MLA soft-decision demapper requires per real symbol $m=3$ additional hardware multipliers to provide the soft bits.

\subsection{Complexity of Soft Deep Neural Network Equalizer}
\label{sec:dnn complexity}

\begin{figure}[b!]
	\begin{center}
		\vspace{-0.6cm}
		\includegraphics[width=3.6in]{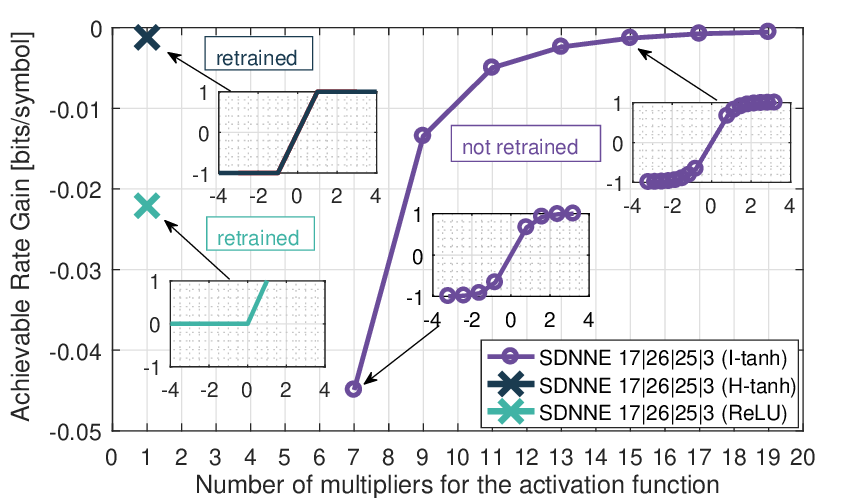}\\	
		\caption{Average achievable rate performance gain in bits/symbol related to the number of required multipliers, when a linear interpolated $\tanh$, a Hard-$\tanh$ or a ReLU activation function is applied. The three subplots illustrate the particular activation functions, where the purple circle indicate the interpolation points.}
		\vspace{-0.3cm}
		\label{GMI_RELU_activation_function}
	\end{center}
\end{figure}

The real-time complexity of a SDNNE is deﬁned by the DNN design and the activation function in use. As described in section~\ref{Principles of Deep Neural Network Equalizers}, the layers interconnect with a weight coefficient per connection. For implementation, those weights require memory space and a hardware multiplier, each, just like the VNLE kernels. In comparison to the VNLE, an additional complexity part comes from the activation functions. The previous chosen $\tanh$ activation function could be in principle implemented with the CORDIC algorithm~\cite{duprat1993cordic}. However, it operates iteratively and thus, slow. A more sleek option is to linearly interpolate the $\tanh$ (I-$\tanh$) and to use a Look-up-table (LUT) or to directly utilize a simpler activation function, e.g., a hard $\tanh$ (H-$\tanh$) or even a Rectified Linear Unit (ReLU) function. The ReLU is a linear function, which becomes nonlinear by clipping. It provides the most simple activation function and requires for implementation only one hardware multiplier and one comparator for clipping. In turn, the H-$\tanh$ function is clipped on two positions, therefore, an additional comparator for clipping is required. The total required number of multipliers for a SDNNE with ReLU or H-$\tanh$ activation function is equal and defined as
\begin{equation}
\text{mul}_{\text{SDNNE-ReLU/H-}\tanh} = \sum_{i=1}^{d-1} s_{i} s_{i+1} + \sum_{i=2}^{d-1} s_i,
\label{mul_SDNNE}
\end{equation}
where $d$ is the number of layers including input and output layer and
where $\boldsymbol{s} = s_1|s_2|\dotsc|s_d$ denotes the SDNNE design, i.e., $s_i$ is the number of units in the $i$-th layer. For instance, the input layer has $i_1=2M+1$ units and the output layer has $i_d=m$ units. The first term is related to the number of weights and the second term to the number of ReLU/H-$\tanh$ activation functions. The complexity of a SDNNE with an I-$\tanh$ activation function depends on the number of interpolation points $K$, namely
\begin{equation}
\text{mul}_\text{SDNNE-Inter.Tanh} = \sum_{i=1}^{d-1} s_{i} s_{i+1} + (K-1) \sum_{i=2}^{d-1} s_i. 
\label{mul_SDNNE}
\end{equation}
The first term is related once more to the number of weights, while the second term is related to the total number of multipliers which are required to generate the particular I-$\tanh$ activation functions. 

\begin{figure}[t!]
	\begin{center}
		\includegraphics[trim = 0.2cm 0 0.2cm 0.0cm,width=3.65in]{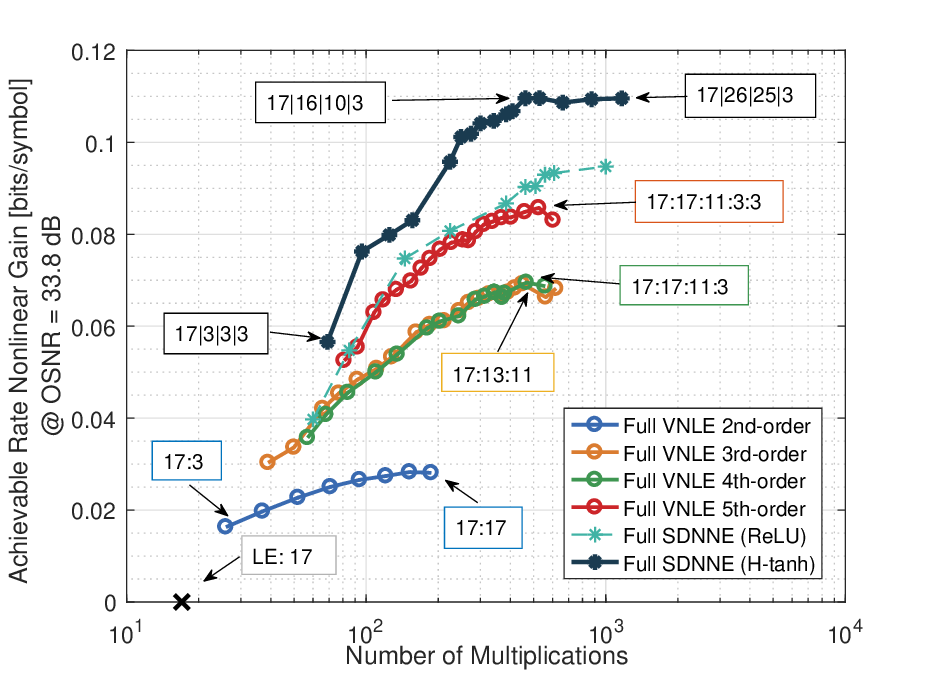}\\
		\caption{Nonlinear compensation gains in bits/symbol related to number of multiplers for a 92-Gbaud 64QAM optical BtB system. The nonlinear equalizer architectures are optimized regarding performance and complexity.  Blue, yellow, green and red represent the VNLE architectures with different orders while the cyan and dark blue represent the different SDNNE architectures with ReLU and H-$\tanh$ activation functions. Selected circled markers are labeled with their architecture.}\label{GMI_multpliers}
	\end{center}
	\vspace{-0.7cm}
\end{figure}

Fig.~\ref{GMI_RELU_activation_function} shows the average achievable rate gain related to the number of multipliers when the previously chosen $\tanh$ activation function is replaced by an I-$\tanh$, a H-$\tanh$ or a ReLU activation function. The average achievable rate gains are evaluated over the particular OSNR captures. The SDNNEs with ReLU and H-$\tanh$ activation functions are retrained, while the SDNNE with I-$\tanh$ activation function 
isn't retrained and hence contains the parameters of the SDNNE with $\tanh$ activation function. It can be observed that the performance of SDNNE with I-$\tanh$ decreases tremendously when less than 16 interpolation points and hence 15 multipliers are used. In contrast, the SDNNE with H-$\tanh$ achieves good results in performance and complexity, while the SDNNE with ReLU loses performance. Therefore, the best trade-off between performance and complexity is achieved with the H-$\tanh$ activation function.

\subsection{Complexity Evaluation} \label{Complexity Evaluation}
\begin{figure}[t!]
	\begin{center}
		\includegraphics[trim = 0.4cm 0 0 0.0cm,width=3.65in]{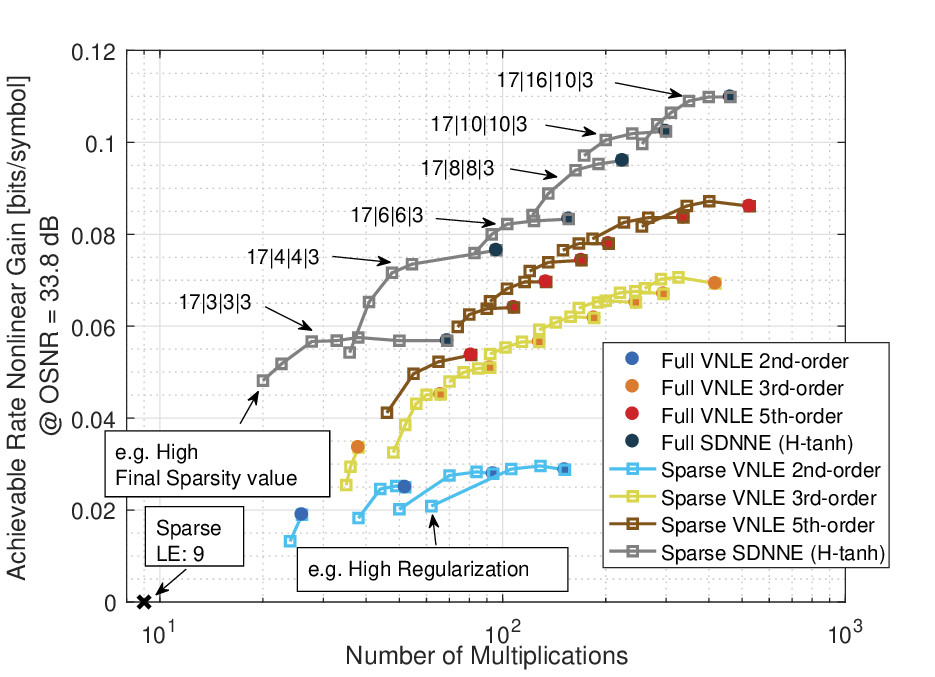}\\	
		\caption{Nonlinear compensation gains in bits/symbol related to number of multipliers. Complexity reduction of VNLE and SDNNE by L1-regularization and weight pruning.}\label{GMI_multpliers_sparse}
	\end{center}
	\vspace{-0.9cm}
\end{figure}
Fig. \ref{GMI_multpliers} depicts the achievable rate gain of the full VNLE and the fully connected SDNNE architectures in relation to the LE versus the utilized number of multipliers at $33.8$dB OSNR (the behaviors are very similar for other OSNR values, see Fig.~\ref{GMI_gain}). All architectures are optimized regarding complexity and achievable rate (e.g. the full VNLE architectures are optimized with respect to the number of linear and nonlinear taps).

The blue, yellow, green and red circled markers indicate individual VNLE architectures up to 5th order. It can be observed that all four curves increase and decrease with the number of multiplications. The overall envelop indicates the optimized trade-off between complexity and performance. In general, it is more efficient to increase the degree of orders than increasing the particular memory sizes. However, as abovementioned kernel orders greater than 5th do not yield any additional gain, but only increased complexity. The best performing VNLE architectures are labeled with their design.

The cyan and dark blue markers represent different SDNNE architectures with ReLU and H-$\tanh$ activation functions, respectively. The cyan line representing the SDNNE with ReLU activation function is dashed due to its limited performance and complexity trade-off and only plotted for completeness. The more interesting architectures, the SDNNE architectures with H-$\tanh$ activation functions, are indicated by the dark blue markers. It can be observed, that the introduced architecture $17|26|25|3$ is very complex but outperforms the VNLE architectures. As shown in Section~\ref{SDNNE}, the 
architecture $17|26|25|3$ assumes that the network has to represent $\sim10^{17}$~activation~patterns for optimal equalization. If this assumption is not applicable, the complexity can be reduced without performance penalty. On the other hand, if complexity is reduced while the required number of activation patterns is essential the SDNNE will exhibit insufficient capacity and will not be able to exactly equalize the signal anymore. Nevertheless, it can still approximate it with an error. In the current setup, by optimizing the SDNNE architecture regarding complexity and achievable rate (same as for the VNLE architectures) the best trade-off is achieved with the design $17|16|10|3$. In comparison to the $17|26|25|3$ architecture no performance loss occurs, which indicates that in this scenario the network has to exhibit a representation capacity of at least~$\sim10^{8}$~activation~patterns.

If less neurons are used, the performance decreases immediately, which indicates that the required number of activation patterns is falling short of and the SDNNE exhibits insufficient representation capacity. Nevertheless, all SDNNE options outperform all optimized full VNLE architectures with equal complexity. 

However, pruning techniques such as the least absolute shrinkage and selection operator LASSO ($\ell1$-norm)~\cite{tibshirani1996regression} for VNLEs or the gradual pruning algorithm~\cite[Sec.~3]{zhu2017prune} for neural networks, have proven to reduce the set of kernels or the number of weights without significant performance loss.

Fig.~\ref{GMI_multpliers_sparse} shows the complexity reductions and the corresponding achievable rate gains of the sparse VNLE and the sparse SDNNE architectures. The filled circled markers represent full architectures, while the particular square markers represent sparse architectures with different regularization values or final sparsity values~\cite{zhu2017prune}, respectively. It can be observed that on average the sparse VNLEs exhibit a $25\%$ lower complexity than the full VNLEs without performance losses. 
The sparse SDNNEs exhibit a pruning factor of $20\%$ for more complex architectures and up to $45\%$ for lower complex architectures. A truncated single weight in a lower complex architecture, with less neurons in the hidden layers, has a larger impact on the overall complexity.

\begin{figure}[t!]
	\begin{center}
			\vspace{-0.2cm}
		\includegraphics[trim = 0.6cm 0 0 0.0cm,width=3.6in]{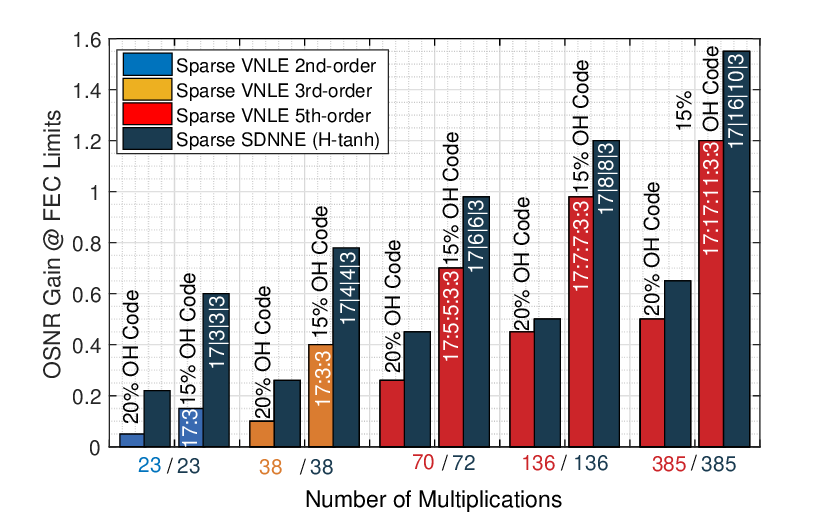}\\
		\caption{OSNR nonlinear compensation gains in dB related to number of multipliers for a 92-Gbaud 64QAM optical BtB system. The labels of the architectures correspond to the $15\%$ as well to the $20\%$ OH code bars. The corresponding FEC requirements are shown in  Fig.~\ref{MI_versus_OSNR}.}\label{GMI_perfromance_complexity}
	\end{center}
	\vspace{-0.9cm}
\end{figure}

Fig.~\ref{GMI_perfromance_complexity} depicts a more detailed representation of the OSNR gains at the assumed FEC limits related to the number of multiplications. The particular gains depend on the assumed FEC overhead and the corresponding FEC limit because of saturation, see Fig.~\ref{MI_versus_OSNR}. Similar or even same complexity architectures are grouped. The sparse VNLE reaches the performance saturation and hence its highest OSNR gain with at least $385$ multipliers. In contrary, the sparse SDNNE achieves similar performance with $136$ multipliers. Its performance saturation and hence the highest OSNR gain is achieved with $385$ multipliers. The VNLE architecture with equal complexity exhibits an OSNR penalty of 0.35dB. In the low complexity regime the sparse VNLE performance decreases significantly more than the SDNNE performance.

\section{Conclusion}
High architectural complexity of Volterra nonlinear equalizers (VNLE) has motivated investigations in nonlinear equalizer alternatives based on deep neural networks (DNN). In optical coherent 92GBd dual polarization 64QAM 950Gb/s back-to-back measurements, where optical and electrical components nonlinearities dominate, the proposed soft DNN equalizers (SDNNE) proved to reflect systematic nonlinearities more accurately than a 5th-order VNLEs. 
They either outperform pruned VNLEs by 0.35dB in OSNR with equal complexity or achieve the same performance with $65\%$ less multipliers and hence lower complexity. In addition, we show that the DNN state-of-the-art cross-entropy cost function for classification problems is equivalent to bitwise equivocation, maximizing an achievable rate. It is therefore optimal for training DSP components acting as soft-demappers in modern communication systems with soft-decision FEC.

\newcommand{\bhat}{\hat{b}}
\newcommand{\sety}{\mathcal{Y}}
\newcommand{\entop}{\mathbb{H}}
\begin{figure*}
	\centering
	\footnotesize
	\begin{tikzpicture}
	\node(mu){$b_i\in\{0,1\}$};
	\node[right = 0.3 of mu,draw](ch){channel};
	\node[right = 0.3 of ch](y){$y_i$};
	\node[right = 0.3 of y,draw](nn){SDNNE};
	\node[right = 0.3 of nn](soft){$\ell_i$};
	\node[right = 0.3 of soft,draw](sigmoid){$\sigma(\cdot)$};
	\node[right = 0.3 of sigmoid](muhat){$\bhat_i\in[0,1]$};
	\draw[-latex](mu)--(ch);
	\draw[-latex](ch)--(y);
	\draw[-latex](y)--(nn);
	\draw[-latex](nn)--(soft);
	\draw[-latex](soft)--(sigmoid);
	\draw[-latex](sigmoid)--(muhat);
	\end{tikzpicture}
	
	\begin{tabular}{cc}
		\toprule
		soft-demapping loss~\eqref{loss function}&logistic regression loss\\\midrule
		$\mathcal{L}(b,\ell)=\begin{cases}
		\log (1+\exp(-\ell)),& \text{if }b=0\\
		\log (1+\exp(\ell)),& \text{if }b=1
		\end{cases}$&$\mathcal{L}(b,\bhat)=\begin{cases}
		\log \frac{1}{\bhat},& \text{if }b=0\\
		\log \frac{1}{1-\bhat},& \text{if }b=1
		\end{cases}$\\
		\bottomrule
	\end{tabular}
	\caption{Bitwise soft-demapping as logistic regression.}
	\label{fig:sigmoid output}
\end{figure*}
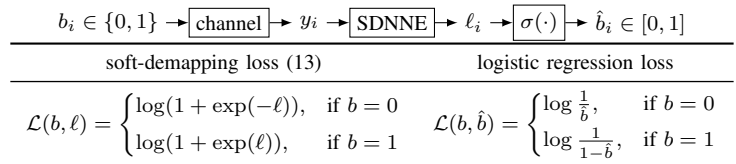

\appendices
\section{Loss Optimality}
\subsection{Soft-Demapping as Logistic Regression}
We now show that our bitwise equivocation loss function $\mathcal{L}(b,\ell)$ defined in \eqref{loss function} that aims to maximize an achievable FEC rate is equivalent to the binary cross-entropy function $\mathcal{L}(b,\bhat)$ applied to the output of a sigmoid activation function $\bhat=\sigma(\ell)$, which is the common approach to binary classification (``logistic regression'') in classic machine learning,  see, e.g., \cite[Sec.~4.3.2]{bishop2006pattern}, \cite[Sec.~8.3.1]{murphy2012machine}, \cite[pp. 142--144]{geron2019hands}. We display the SDNNE and the two loss functions under consideration in Fig.~\ref{fig:sigmoid output}. We next show that the two loss functions are equivalent. The sigmoid activation function~\cite[Sec.~2.4]{bishop2006pattern} is defined as
\begin{align}
\bhat=\frac{1}{1+\exp(-\ell)}.\label{eq:sigmoid}
\end{align}
We now have
\begin{align}
\mathcal{L}(b,\bhat)&=\begin{cases}
\log \frac{1}{\bhat},& b=0\\
\log \frac{1}{1-\bhat},& b=1
\end{cases}\\
&\overset{\eqref{eq:sigmoid}}{=}\begin{cases}
\log (1+\exp(-\ell)),& b=0\\
\log (1+\exp(\ell)),& b=1
\end{cases}\label{eq:sigmoid in xe}\\
&=\log_2 (1+\exp(-(1-2b)\ell))\label{eq:loss equivalence}\\
&=\mathcal{L}(b,\ell)
\end{align}
which shows the equivalence of the loss functions.

\subsection{Optimal Demapping with SDNNE}

For a trained SDNNE, the SDNNE outputs $\ell$ and $\bhat$ are deterministic functions of the SDNNE input $y$. We define
\begin{align}
q(0,y) = \bhat(y),\quad q(1,y)=1-\bhat(y).
\end{align}
Note that since $0\leq\bhat(y)\leq 1$, for each $y$, $q(\cdot, y)$ defines a distribution on $\{0,1\}$. We can now write the loss as
\begin{align}
\mathcal{L}(b, \bhat(y))=-\log_2 q(b, y).
\end{align}
For a sufficiently large number $n$ of samples, the average loss is the cost that is minimized by training. We next derive an information-theoretic interpretation of the cost. To this end, we assume the channel output $y$ takes values in a finite set $\mathcal{Y}$ (Note that this assumption is in line with a practical setting, where $y$ is discrete because of analog-to-digital conversion with finite resolution).  For $n$ samples, the cost is
\begin{align}
& \frac{1}{n}  \sum_{i=1}^{n}[-\log_2 q(b_i, y_i)]  \\ &=  \frac{1}{n} \sum_{b \in \{0,1\}} \sum_{y \in \mathcal{Y}} \sum_{i: b_i=b, y_i = y} [-\log_2 q(b, y)] \\
&= \sum_{b \in \{0,1\}} \sum_{y \in \mathcal{Y}} [-\log_2 q(b, y)]\underbrace{\frac{\sum_{i: b_i=b, y_i = y}1}{n}}_{(\star)},
\end{align} 
where the term $(\star)$ defines a joint distribution on $\{0,1\} \times \mathcal{Y}$. Let's introduce the random variable (RV) $B$ for the channel input and the RV $Y$ for the channel output observed by the SDNNE and define 
\begin{equation}
P_{BY}(b,y) = \frac{\sum_{i: b_i=b, y_i = y}1}{n}.
\end{equation}
Then, we have 
\begin{align}
&\frac{1}{n} \sum_{i=1}^{n} [-\log_2 q(b_i, y_i)] = \sum_{b \in \{0,1\}} \sum_{y \in \mathcal{Y}} P_{BY}(b,y) [-\log_2 q(b, y)]\\
&= \sum_{y \in \mathcal{Y}} P_Y(y) \sum_{b \in \{0,1\}} P_{B|Y}(b|y)  [-\log_2 q(b, y)]\label{eq:bayes}\\
&\geq \sum_{y\in\sety}P_Y(y)\sum_{b\in\{0,1\}}P_{B|Y}(b|y)[-\log_2 P_{B|Y}(b|y)]\label{eq:information inequality}\\
&=\entop(B|Y)
\end{align}
where \eqref{eq:bayes} follows by Bayes' rule and \eqref{eq:information inequality} by the information inequality \cite[Theorem~2.6.3]{cover2006elements} (recall that $q(\cdot,y)$ defines a distribution on $\{0,1\}$ for each $y$). We have equality in \eqref{eq:information inequality}
if and only if
\begin{align}
\bhat(y)=q(0,y)&=P_{B|Y}(0|y)\\
1-\bhat(y)=q(1,y)&=P_{B|Y}(1|y).
\end{align}
We next characterize the optimal $\ell$. For $b=1$, we have
\begin{align}
&\frac{1}{1-\bhat(y)}=1+\exp[\ell(y)]=\frac{1}{P_{B|Y}(1|y)}\\
&\Rightarrow\exp[\ell(y)]=1-\frac{1}{P_{B|Y}(1|y)}=\frac{P_{B|Y}(0|y)}{P_{B|Y}(1|y)}\\
&\Rightarrow\ell(y)=\log\frac{P_{B|Y}(0|y)}{P_{B|Y}(1|y)}.
\end{align}
(The same result is obtained when considering $b=0$). We conclude that with the considered loss functions, the SDNNE learns via $\bhat$ the a-posteriori probability (APP) distribution $P_{B|Y}$ and via $\ell$, it learns the logarithmic APP ratio $\log\frac{P_{B|Y}(0|y)}{P_{B|Y}(1|y)}$, thereby achieving the minimum cost of $\entop(B|Y)$.

\ifCLASSOPTIONcaptionsoff
  \newpage
\fi

\bibliographystyle{IEEEtran}
\bibliography{IEEEabrv,Bibliography,Bibliography-gb}

\vfill
\end{document}